# ELECTRON EMISSION AT VERY LOW ELECTRON IMPACT ENERGY: EXPERIMENTAL AND MONTE-CARLO RESULTS *

M. Belhaj, J. Roupie, ONERA The French Aerospace Lab, Toulouse, France
O. Jbara, LISM, UFR Sciences, Univeristy of Reims, France
J. Puech, N. Balcon, D. Payan CNES, Toulouse, France

*Abstract*

The behaviour of electron emission under electron impact at very low energy is of great importance in many applications such as high energy physics, satellites, nuclear reactors, etc. However the question of the total electron reflectivity is still in discussion. Our experimental and theoretical studies show that the total reflectivity at very low energy is far from being an obvious fact. Moreover, our results show that the yield is close to zero and not equal to one for low energy incident electron.

## INTRODUCTION

In recent works, it has been shown that the behaviour of electron emission under electron impact at very low energy (few eV) could considerably affect the multipactor Radio Frequence (RF) discharge susceptibility diagrams (more particularly the resonant mode 1) [1]. Unfortunately, published results on the total electron emission yield (TEEY) at very low incident energy are very scarce and when the data is available the results are sometimes in disagreement. The TEEY is the ratio of the number of emitted electrons (secondary and backscattered) to the number of incident electrons. According to the literature, the evolution of the yield when the incident electron energy decreases to zero is still unclear. In some cases, the yield decreases to zero [2,3] whereas in other cases, it grows to one [4,5]. At low energies, the electron trajectories are highly sensitive to the slightest electric or magnetic disturbance (earth magnetic field or stray electric field in the vacuum chamber). This makes the measurement of the yield very difficult. An experimental facility at ONERA/DESP was developed and specially designed to assess the electron emission properties at low incident energies (up to 2 keV and down to a few eV). Previous realistic PIC simulations of such a facility indicated that some methods used for electron emission yield measurements can become inaccurate for low energy incident electrons, even without any magnetic stray field [6]. Here, we present results of SEY obtained on an optimized experimental setup which permits to clearly have access to the range of low energy incident electrons. The results are compared experimental results and to Monte-Carlo results.

## EXPERIMENTAL SETUP

The experimental setup is shown in Figure 1. Cryogenic pump associated to oil-free molecular-diaphragm pumps allow the system to be maintained at vacuum level down to $3.10^{-7}$ Torr.

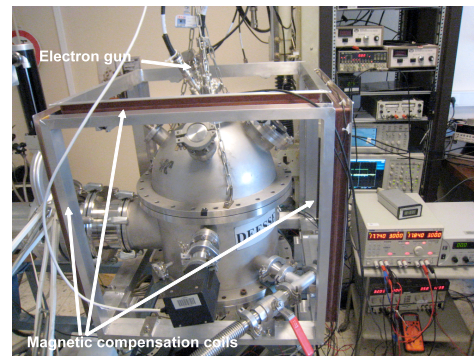

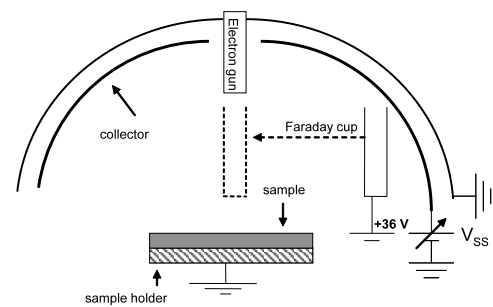

Figure 1: – Picture (up) and schematic view (down) of the ONERA secondary electron experimental facility DEESSE

A hemispherical electron-collecting electrode (collector) faces the sample surface. The facility is equipped with six magnetic coils, used to compensate the static magnetic field (i.e. earth magnetic field) into the vacuum chamber. The sample holder and the collector can be independently biased to choose the desired potential. Sample current is monitored using 350 Mhz TDS5034B oscilloscope connected to a Femto-DHPCA-100 high speed and low noise current amplifier. The electron beam incidence is set normal to the sample surface. A low energy (1eV-2000 eV) ELG2 Kimball Physics electron gun with a microsecond electron beam pulsing capacity was used as the electron source. The incident charge per pulse, $\Delta Q_i$ is measured using a Faraday cup connected to oscilloscope throughout a second Femto-DHPCA-100 current amplifier. The total secondary emission electron yield can be obtained from relation (1).

$$\sigma = \frac{\Delta Q_i - \Delta Q_S}{\Delta Q_i} \qquad (1)$$

$\Delta Q_S$ is the flowed charge from the sample to the ground. The incident current $\Delta Q_i$ is typically about 0.5 pC.

## EXPERIMENTAL RESULTS

The electron emission yields measured on several "technical materials" are shown in Figure 2. The lower investigated incident energy is about 3 eV. In contrast to previous reported results [4,5], no rapid increase of the yield to one at very low energy was observed. Interesting observed feature on the yields is a local maximum observed typically around 4 to 7 eV. Bronshtein and Fraiman [7] as well as Khan et al [3] had observed and reported similar structures.

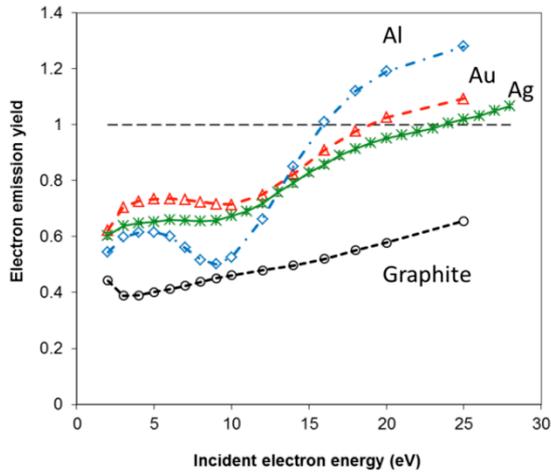

Figure 2: – Total electron emission yield of technical materials (Aluminium, Gold, Silver and Graphite). It should be noticed that all the samples were long time exposed to ambient atmosphere.

## SIMULATION RESULTS

The simulation tool OSMOSEE (Onera Simulation Model Of Secondary Electron Emission) is a Monte-Carlo simulation code dedicated to electron emission in sub-keV energy range. OSMOSEE is described in details in Ref [8]. Briefly, the followed approach consists in handling the interactions within the solid (elastic and inelastic). By ionization, electrons are moved within the target material and are treated like the primary electrons. The whole electronic cascade is followed and focus is given on electrons leaving the target: secondary and backscattered electrons. Simulation results lead in particular to the electron emission yield as a function of incident energy or incident angle, to the emitted electron energy as a function of incident energy and to the angular distribution of emitted electrons. For each result, we have access to primary and secondary electron contributions.

The interactions implemented in OSMOSEE are the following:

- Electron-atom interaction (elastic interaction)
- Electron-free electron (inelastic interaction)
- Electron-volume plasmon interaction (collective effect)
- Electron-surface plasmon interaction (collective effect)
- Single electron-electron interaction
- Electron-bound electron (inelastic interaction)
- Electron-surface interaction, treated with a quantum approach of a realistic potential barrier for transmission probability. The modelling of the barrier takes into account the charge image effect.

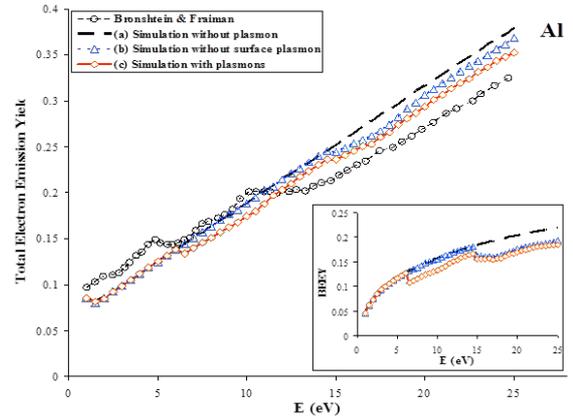

Figure 3: – Total Electron Emission Yield simulated in three cases: (a) without plasmon, (b) with volume plasmons but no surface plasmon, (c) with volume and surface plasmons; compared to experimental TEEY of Bronshtein and Fraiman [7]. The insert represents the Backscattered Electron Emission Yield η calculated with same simulation cases as TEEY.

Simulation results performed on Al are compared to experimental data of Bronshtein and Fraiman [7] obtained on very pure Al. A complete explanation of the observed structures on the TEEY is given in [8]. The good agreement near zero energy ends the ambiguities concerning the emission yield that tends to 1 as the energy tends to zero. At few eV, the drop of the electron emission is explained by the limit angle of surface attack and the competition between elastic escape probability and inelastic interaction probability. The exponential barrier model of the quantum reflection probability used in OSMOSEE simulation does not show any noticeable

increase until few tenths eV, contrary to the rough square barrier model [9]

## DISCUSSION AND CONCLUSION

Careful experimental and theoretical investigations showed that the SEY does not tend to unity for very low primary electron energy, as sometimes systematically reported in the literature. The fact that the SEY could tend or not to 1 for very low primary electron energy has a direct impact on the shape of the resonant mode 1 of the Multipactor susceptibility diagram. As a matter of fact, a shift of the left boundary of this zone has been observed as results of the simulation campaign. As it is in contradiction with agreed picture of the Multipactor susceptibility diagram, the conclusion was that one has to be very careful about the shape of the secondary emission yield curve, more particularly at low incident electron energy level. The SEY curve measured on technical materials set-up confirms the fact that the situation when the SEY tends to 1 for very low primary electron energy is not a general rule. Moreover, the OSMOSSE Monte-Carlo modelling specially developed for simulation of the secondary electron emission yield at very low energy and taking into account a realistic surface potential barrier does not show any increase of the yield to 1 at very low energy. The results presented here are in agreement with the TEEY measurements performed by Bronshtein and Fraiman [7] and presented by Andronov et al [10] at this conference (e-cloud 12).

*Work supported by the CNES
Mohamed.Belhaj@onera.fr